\documentclass[twocolumn,showpacs,preprintnumbers,amsmath,amssymb]{revtex4}
\usepackage{graphicx}
\usepackage{dcolumn}
\usepackage{bm}
\newif\ifappendixon
\begin{document}

\preprint{APS/123-QED}

\title{Slow-light solitons: influence of relaxation}

\author{A.V. Rybin }
\affiliation{Department of Physics, University of Jyv\"askyl\"a PO
Box 35, FIN-40351
\\ Jyv{\"a}skyl{\"a}, Finland\\ St Petersburg University of Information Technologies,
Mechanics and Optics, Kronwerkskii ave 49, 197101, St Petersburg,
Russia } \email{andrei.rybin@phys.jyu.fi}

\author{ I.P. Vadeiko}
\affiliation{School of Physics and Astronomy, University of St
Andrews, North Haugh, St Andrews, KY16 9SS, Scotland}
\email{iv3@st-andrews.ac.uk}

\author{ A. R. Bishop}
\affiliation{Theoretical Division and Center for Nonlinear Studies,
Los Alamos National Laboratory, Los Alamos, New Mexico 87545, USA}
\email{arb@lanl.gov}

\date{ September 19, 2006}

\begin{abstract}
We have applied the transformation of the slow light equations to
Liouville theory that we developed in our previous work, to study
the influence of relaxation on the soliton dynamics. We solved the
problem of the soliton dynamics in the presence of relaxation and
found that the spontaneous emission from the upper atomic level is
strongly suppressed. Our solution proves that the spatial shape of
the soliton is well preserved even if the relaxation time is much
shorter than the soliton time length. This fact is of great
importance for applications of the slow-light soliton concept in
optical information processing. We also demonstrate that the
relaxation plays a role of resistance to the soliton motion and
slows the soliton down even if the controlling field is constant.

\end{abstract}
\maketitle

\begin{figure}
\includegraphics[width=50mm]{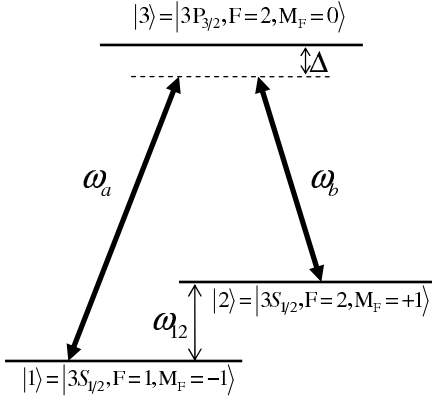}
\caption{\label{fig:spec1} The  $\Lambda$-scheme for  sodium atoms.}
\end{figure}

The development of modern methods of optical signal manipulation and
control opens wide perspectives for application of the light in
classical and quantum computation. One of such methods of optical
signal processing is based on the effect of electromagnetically
induced transparency (EIT)~\cite{Fleischhauer:2005}, which allows
for slowing the light down by many orders of magnitude and even
bringing it to a complete halt~\cite{Hau:1999, Liu:2001,
Phillips:2001}. The major advantage of this methods is that the
velocity of the optical signal is effectively controlled by an
auxiliary laser~\cite{Fleischhauer:2000, Matsko:2001, Bajcsy:2003}.
In the linear regime of operation when the intensity of the
controlling field is significantly larger than the intensity of the
signal, there are some important constraints imposed on the
parameters of the signal~\cite{Fleischhauer:2005}. These constraints
result from the presence of strong optical relaxation in the medium.
The medium is typically a gas of alkali atoms whose electronic
structure relevant to the EIT effect can be schematically described
by the three-level $\Lambda$ model (see Fig.~\ref{fig:spec1}). The
relaxation results from spontaneous transitions of the atoms from
the excited upper energy level $|3\rangle$. Usually, such
transitions may occur not only to the levels $|1\rangle$,$|2\rangle$
but also to other lower levels, which are not included into
consideration for the sake of simplicity. In any case, the
relaxation destroys optical coherence of the signal and therefore
must be accounted for.

The atom-field interaction in the system with EIT is substantially
nonlinear. In our previous works we provided a nonlinear solution of
the EIT problem called the slow-light soliton~\cite{Rybin:2005}. We
also demonstrated that the dynamics of slow-light solitons strongly
depends on the form of the controlling field and the solitons can be
effectively manipulated concerning their  space-time
dynamics~\cite{ryb7, ryb9}. In our present work we study the
dynamics of the solitons in the presence of the relaxation. We note
that in the framework of the nonlinear theory there might exist
different approaches to effectively suppress the influence of
relaxation. Here, we demonstrate that appropriately choosing the
parameters of the controlling field we can reduce the influence of
the relaxation on the slow-light solitons by several orders of
magnitude. Our results are found to be in a good agreement with the
experiments on EIT.

Within the slowly varying amplitude and phase approximation (SVEPA)
the  Maxwell equations for the Rabi-frequencies   are reduced to the
the well-known
 equations governing the dynamics of the atom-field system, viz.
\begin{eqnarray}\label{Maxwell_r}
    \partial_\zeta \Omega_a = i \nu_0 \,\psi_3 \psi_1^*, \;
    \partial_\zeta \Omega_b = i \nu_0 \,\psi_3 \psi_2^*.\quad
\end{eqnarray}
Here $\zeta=z/c, \tau=t-z/c$. The Schrodinger equation for the
amplitudes $\psi_{1,2,3}$ of atomic wave function reads
\begin{eqnarray}\label{Schrod}
    \partial_\tau \psi_1 &=& \frac i2 \Omega_a^*\,\psi_3;\nonumber\\
    \partial_\tau \psi_2 &=& \frac i2 \Omega_b^*\,\psi_3;\\
    \partial_\tau \psi_3 &=& -(i\,\Delta+\frac\gamma2)\,\psi_3 +
    \frac i2(\Omega_a\psi_1+\Omega_b \psi_2).\nonumber
\end{eqnarray}
By excluding the amplitudes of the lower levels the system of
Eqs.(\ref{Maxwell_r}), (\ref{Schrod}) can be transformed to the
following form
\begin{eqnarray}
   & \frac1{\psi_3^*}\partial_{\tau}\frac1{\psi_3}\partial_{\zeta}\Omega_a
  = \frac{\nu_0}{2}\Omega_a;\label{MB_eqs_1}\\
   & \frac1{\psi_3^*}\partial_{\tau}\frac1{\psi_3}\partial_{\zeta}\Omega_b
  = \frac{\nu_0}{2}\Omega_b;\label{MB_eqs_2}\\
   & \partial_\tau |\psi_3|^2 =-\gamma\,|\psi_3|^2  -\frac1{2\nu_0}\partial_{\zeta}
    (|\Omega_a|^2+|\Omega_b|^2).\label{MB_eqs_3}\\
   & \partial_\tau \varphi_3 = -\Delta +\frac1{2\nu_0\,|\psi_3|^2}
   (|\Omega_a|^2\,\partial_{\zeta}\varphi_a+
   |\Omega_b|^2\,\partial_{\zeta}\varphi_b).\label{MB_eqs_4}
\end{eqnarray}
Here, $\varphi_{a,b,3}$ are respectively the phases of the fields
$\Omega_{a,b}$ and the amplitude of the excited state $\psi_3$.
Notice that the equations Eqs.~(\ref{MB_eqs_1}),(\ref{MB_eqs_2}) are
the wave equations for the fields in curvilinear coordinates with
the metric dependent on the amplitude $\psi_3$.

In the typical dynamics scenario, the atoms are initially prepared
on the lowest energy level $|1\rangle$, the probe field is absent
while the controlling field is constant. Notice that the state
$|1\rangle$ is a dark state for the controlling field. Thus, the
initial state of the system  reads
\begin{equation}\label{init_fields}
    \Omega_a(0,\zeta)=0,\; \Omega_b(0,\zeta)=\Omega_0,\;
    |\psi_{at}(0,\zeta)\rangle=|1\rangle.
\end{equation}

In fact, this is the  simplest solution of the equations
Eqs.(\ref{MB_eqs_1}),(\ref{MB_eqs_2}),(\ref{MB_eqs_3}),(\ref{MB_eqs_4}).

\section{\label{sec2} Relaxation}
From the results of linear and nonlinear theories of
electromagnetically induced transparency (EIT) ~\cite{Harris:1997,
Grobe:1994, Dey:2003, Dutton:2004, Rybin:2004, ryb7} we infered a
physically plausible assumption that the population of the excited
state $|3\rangle$ is proportional to the amplitude of the field in
the probe channel, which we denote as $\Omega_a$. In our previous
works we demonstrated that a slow-light soliton in the system
without relaxation is obtained by taking
$\psi_3=-\frac1{2|\lambda-\Delta|}\Omega_a$. Here, $\lambda$ is an
arbitrary parameter limited from below by the condition
$|\psi_3|\leq1$. In order to take into account the relaxation we
propose a more general relation between the amplitude of atomic
upper level $|3\rangle$ and the amplitude $\Omega_a$ of the
slow-light soliton. The population of the upper level although very
small should remain nonzero in order to preserve necessary coupling
between the probe and controlling field. At the same time, the
relaxation should manifest itself in the form of some effective
damping of the probe field. Based on such consideration we found the
following relation between the amplitudes:
\begin{equation}\label{psi3_Omegaa}
    e^{\alpha(\tau)}\psi_3=-\frac{1}{2|\lambda-\Delta|}\Omega_a,
\end{equation}
where $\alpha(\tau)$ is a first order correction to the exact
slow-light soliton solution. It will be shown below that
$\alpha(\tau)$ is always negative. Let us introduce convenient
notations
$$|\Omega_a|\equiv e^{-\rho},\, \Omega_b\equiv\eta,
\;\tilde\rho \equiv \rho+\alpha,\;
    k=\frac{\nu_0}{8|\lambda-\Delta|^2}.$$
The first three equations of the system Eq.(\ref{MB_eqs}) take then
the following form:
\begin{eqnarray}
    &\label{Liouv1}\partial_{\tau}\alpha\;  \partial_{\zeta}\tilde\rho +
    \partial_{\zeta \tau}\tilde\rho=-k\,e^{-2\tilde\rho}\\
    &\label{Liouv2}
    \partial_{\zeta \tau}\eta+\partial_{\tau} \tilde\rho\;
    \partial_\zeta \eta=k\,e^{-2\tilde\rho}\,\eta,\\
    & \label{Liouv3}
    4k(\partial_{\tau} +\gamma)\;
    e^{-2\tilde\rho}=-\partial_{\zeta}\left({ \eta^2 +
    e^{-2\rho}}\right).
\end{eqnarray}
The last equation for the phase $\varphi_3$ can easily be integrated
after we solved the first three.

Assuming that the effective relaxation described by $\alpha$ varies
slowly in time $\tau$ and therefore neglecting the first term in
Eq.(\ref{Liouv1}) we can find a solution of the system of equations
Eqs.(\ref{Liouv1}), (\ref{Liouv2}), (\ref{Liouv3}). We will provide
a rigorous description of this approximation after we construct the
solution in the next section.

\section{\label{sec3} Solution}
We  assume that $\partial_{\tau}\alpha$ is small as compared to the
r.h.s. of the equation Eq.(\ref{Liouv1}), and neglect the first term
in the l.h.s of this equation. The equation then transforms into the
well known Liouville equation, viz.
\begin{equation}\label{Liouville_exact}
    \partial_{\zeta \bar\tau}\tilde\rho=-k\,e^{-2\tilde\rho},
\end{equation}
whose  general solution is readily available:
\begin{equation}\label{Liouv_sol}
\rho=-\frac12 \log\left[{ \frac{\frac 1k e^{2\alpha}
\partial_\zeta\,A_+(\zeta)
\partial_{\tau}\,A_-(\tau) }
{(1- A_+ A_-)^2}}\right].
\end{equation}
Here $A_+(\zeta),A_-(\tau)$ are arbitrary functions. To obtain the
solution of the whole system including Eqs.(\ref{Liouv2}) and
(\ref{Liouv3}) we specify these arbitrary functions as follows:
\begin{eqnarray}\label{sol1}
& A_+(\zeta)=-\exp[-8\varepsilon_0 k \zeta],\\
& A_-(\tau)=\exp\left[{2\varepsilon_0 \int
\frac{e^{2\alpha(\tau)}}{p(\tau)^2+1}d\tau}\right],\\
&\label{sol4}\eta=-2p\,
\partial_\tau\rho+2\partial_\tau p-2p\,\partial_\tau\alpha ,\\
\nonumber\\
&\label{sol5}\partial_\tau\alpha(\tau)=-\frac{\gamma/2}{p(\tau)^2+1}.
\end{eqnarray}
$p(\tau)$ is an arbitrary function describing the controlling field
$\Omega(\tau)$, while $\lambda=i\,\varepsilon_0$. We note from
Eq.(\ref{sol5}) that the correction $\alpha$ to exact solution
without relaxation obtained in our previous papers vanishes for
$\gamma=0$ as expected.

So, the fields read
\begin{eqnarray}\label{fields1}
&\Omega_a(\tau,\zeta)=\frac{2 \varepsilon_0
e^{2\alpha}}{\sqrt{p(\tau)^2+1}}
\mathrm{sech}(\varphi),\\
& \label{fields2}\Omega_b(\tau,\zeta)=-\frac{2
\varepsilon_0\,p(\tau)e^{2\alpha}}{p(\tau)^2+1} \tanh(\varphi)+
\frac{2\partial_\tau p(\tau)-\gamma p(\tau)}{p(\tau)^2+1}\nonumber
\end{eqnarray}
with the phase
\begin{equation}\label{soliton_phase}
 \varphi=-4k\,\varepsilon_0\,(\zeta-\zeta_0)+\varepsilon_0
 \frac{1-e^{2\alpha(\tau)}}\gamma.
\end{equation}
We chose $\alpha(0)=0$. From the solution for the phase of
slow-light soliton Eq.(\ref{soliton_phase}) we obtain its velocity:
\begin{equation}\label{sol_velocity}
    v_g=\frac1{4k}\frac{e^{2\alpha}}{p(\tau)^2+1}.
\end{equation}

\section{\label{sec4}  Validity of the approximation}
Now we will analyze the validity of our approximation of small
$\partial_\tau\alpha$. The first term of Eq.(\ref{Liouv1}) can be
neglected if it is much smaller than $k\,e^{-2\tilde\rho}$. Hence we
find the following condition on $\alpha$:
\begin{equation}\label{approx}
    \left|{\partial_{\tau}\alpha}\right|\ll
   \left|{
   \frac{8\varepsilon_0 e^{2\alpha}}{(p^2+1)\sinh(2\varphi)}}\right|.
\end{equation}
This condition can be simplified with the help of Eq.(\ref{sol5})
and we obtain
\begin{equation}\label{condit}
    \frac\gamma{16 \varepsilon_0} \ll
    \frac{e^{2\alpha}}{|\sinh(2\varphi)|}.
\end{equation}
It is always fulfilled at the maximum of slow-light soliton, because
$\varphi=0$ there. The position of the maximum of the soliton is
given by the following function of retarded time $\tau$:
\begin{equation}\label{soliton_max}
    \zeta_c(\tau)=\frac{1-e^{2\alpha(\tau)}}{4k\gamma}+\zeta_0,
\end{equation}
where we chose $\zeta_c(0)=\zeta_0$. Hence, the phase of the soliton
can be rewritten as
\begin{equation}\label{soliton_phase2}
\varphi=-4k\,\varepsilon_0\,(\zeta- \zeta_c(\tau)).
\end{equation}
From equation Eq.(\ref{condit}) we can find the spatial window of
validity  of our approximation with respect to the center of
slow-light soliton $\Delta \zeta(\tau)=\zeta- \zeta_c(\tau)$ where
our soliton provides correct description of the pulse shape. It
reads
\begin{equation}\label{delta_z}
    |\sinh(-8k\,\varepsilon_0\; \Delta \zeta)| \ll
    \frac{16 \varepsilon_0}\gamma e^{2\alpha}.
\end{equation}
At the initial moment of time, for $\gamma<16\varepsilon_0$ we find
$k\,\varepsilon_0\, \Delta \zeta_0\approx
\ln\left({\frac{2\varepsilon_0}\gamma}\right)$. We will denote the
full-width at half-maximum of slow-light soliton by $w_s\approx
0.66/(k|\varepsilon_0|)$. To keep our solution valid at least within
the full-width at half-maximum of the soliton, i.e. $w_s=2\Delta
\zeta_0$, the parameter $\varepsilon_0$ should obey the following
condition
\begin{equation}\label{epsilon_gamma}
    \varepsilon_0\geq 0.7\gamma
\end{equation}

Notice that $\alpha$ is a negative monotonically decreasing function
of $\tau$. Therefore, the validity window closes down with the
retarded time $\tau$. However, the group velocity of the slow-light
soliton approaches zero as $\tau$ increases and the soliton slows
down to full stop (see Eq.(\ref{sol_velocity})). The distance that
the soliton travels until full stop is
\begin{equation}\label{distance}
    {\cal L}\equiv\zeta_c(\infty)-\zeta_0\leq\frac1{4k\gamma}.
\end{equation}
This formula clearly shows that the maximum distance that our
soliton can travel in the medium is limited by the magnitude of the
relaxation constant. As stronger the relaxation is, as smaller the
distance that the soliton can propagate in the medium.

\section{\label{sec5} Comparison with experiments}
We discussed in our previous papers that the controlling field
generated by an auxiliary laser on entrance into the medium supports
the propagation of the slow-light soliton. An appropriately chosen
modulation of the controlling field ensures the nonlinear coupling
between the soliton and atoms, while the nonlinear coupling in its
turn preserves the secant shape of the soliton. However, in the
ideal case when the relaxation is absent, the modulation of the
controlling field has the same time length as the signal in the
channel a. After the soliton has been generated, the auxiliary laser
continues emitting a constant laser beam, which supports the
propagation of the soliton with a constant velocity. We demonstrated
that this ideal situation when the amplitude of the soliton remains
constant corresponds to a constant function $p(\tau)=p_0$. If the
relaxation is not zero, the amplitude and the velocity of the
soliton will be decaying in time $\tau$ according to
Eq.(\ref{fields1}). The asymptotic value of the background field
past the modulation supporting the soliton can be found from
$\Omega_b(\tau,\zeta)$ by taking the limit $\zeta\rightarrow\infty$
at the zero moment of time:
\begin{equation}\label{bg_field}
    \Omega_0=\frac{(2|\varepsilon_0|-\gamma) p_0}{p_0^2+1}.
\end{equation}
Assuming that $\gamma \gg \Omega_0$ we find
\begin{equation}\label{p_0}
    p_0=\frac{2|\varepsilon_0|-\gamma}{2\Omega_0}+
    \sqrt{\left({\frac{2|\varepsilon_0|-\gamma}{2\Omega_0}}
    \right)^2-1}\simeq\frac{2|\varepsilon_0|-\gamma}{\Omega_0}.
\end{equation}
In the absence of the relaxation, the time length of the soliton
pulse on entrance into the medium is determined by the phase
$\varphi_0(t)=\varepsilon_0\,t/(p_0^2+1)$. Hence we obtain a second
condition on two arbitrary parameters $p_0$ and $\varepsilon_0$ in
terms of well-defined experimental parameters $\Omega_0$ and $t_p$:
\begin{equation}\label{time_width}
    \mathrm{sech}\left({\frac{|\varepsilon_0|}{p_0^2+1}\;t_p/2}\right)=0.5
\end{equation}
Choosing $|\varepsilon_0|=\gamma$ we verify that the conditions
Eqs.(\ref{epsilon_gamma}), (\ref{bg_field}), (\ref{time_width}) are
well satisfied for experimentally feasible values of $t_p\sim 1\mu
s$ and $\Omega_0\sim 10\, \mathrm{MHz}$.

\begin{figure}
\includegraphics[width=80mm]{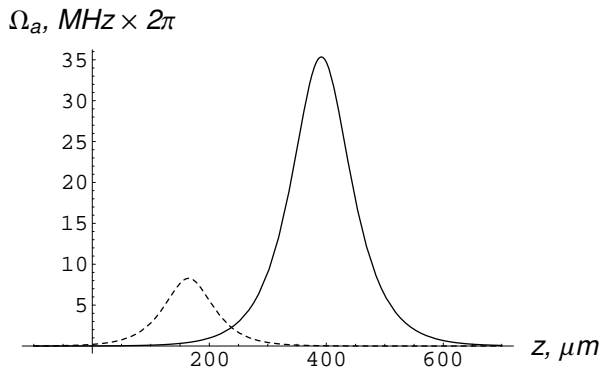}
\caption{\label{fig:spec2} Slow-light soliton decay due to the
relaxation. The solid line corresponds to a reference soliton
propagating in the system without relaxation, and the dashed line
represents the pulse in the presence of relaxation.}
\end{figure}

Now we will compare our solution with the experimental results
reported for sodium atoms~\cite{Hau:1999}. The parameters have the
following values: $\gamma=6.3\times10^7\; \mathrm{rad}\,
\mathrm{s}^{-1}$, $t_p=2.5 \mu s$, and $\Omega_0=0.56 \gamma$. We
solve Eqs.(\ref{bg_field}), (\ref{time_width}) and find
$p_0\simeq18.4$, $\varepsilon_0\simeq5.7 \gamma$. We can also
calculate the reduction in the strength of optical relaxation
influencing the dynamics of the slow-light soliton. According to our
solution Eq.(\ref{fields1}), the amplitude of the soliton decays
with the rate $2\alpha$. Therefore, we can introduce an effective
relaxation constant defined as follows (see Eq.(\ref{sol5}))
\begin{equation}\label{effective_gamma}
    \gamma^*=\frac{\gamma}{p_0^2+1}.
\end{equation}
In the case considered above, $\gamma^*\simeq\gamma/340$. We note
that the effective value of the relaxation constant $\gamma$ is
significantly lower for the soliton than for an arbitrary pulse. So,
the spontaneous emission is greatly suppressed for soliton pulses
due to the nonlinear interaction with the medium. The effective
optical relaxation time $\tau_{rel}^*$ is larger than the pulse
length and approximately equals $2.2\, t_p$. For the pulse delay
$\Delta t=7.05\, \mu s$ reported by Hau~\cite{Hau:1999} we
calculated the decay of the soliton pulse amplitude at the maximum
and compared it with a reference pulse $\Omega_a^*$ propagating in
the system without relaxation. We obtained
$$\frac{\Omega_a(\Delta \tau+t_p/2,\zeta_c)}
{\Omega_a^*} \simeq 0.2$$ The reference pulse $\Omega_a^*$ is
modeled traveling under the same conditions in the system without
relaxation. We added a half of the pulse time width, because it
started interacting with the medium approximately $t_p/2$
microseconds earlier than the maximum entered inside
Fig.~\ref{fig:spec1}. This result agrees very well with the
measurements reported by Hau. We also note that the distance that
the slow-light soliton propagated in the medium during the time
$\Delta \tau+t_p$ is approximately equal $200 \mu m$, which is of
order of the atomic cloud in the experiment~\cite{Hau:1999}. We need
to emphasize that in the presence of the relaxation the velocity of
the soliton is not a constant any more like in the ideal case (see
Eq.(\ref{sol_velocity}). We calculated that the average value of the
velocity is approximately $22 m/s$ compared to experimentally
measured $32 m/s$ for the Gaussian pulse.

A more important feature of the dynamics is in the fact that the
soliton spatial form does not depend on the relaxation at all. The
soliton shape remains unchanged along the propagation inside the
medium (see Fig.~\ref{fig:spec1}). This fact plays a crucial role
for storage and retrieval of optical information in potential
applications.

\section{\label{sec10} Discussion}
The results of our work remain valid beyond the constraints of the
transparency window defined in the linear theory. We demonstrated
that due to strong nonlinear interaction between the probe and
controlling fields it is possible to preserve the spacial shape of
the optical signals even in the presence of strong optical
relaxation. The comparison of our theory and experimental results
~\cite{Hau:1999, Liu:2001} shows a very good agreement. We provided
rigorous analytical estimates for the largest distance that the
slow-light soliton can propagate in the medium with relaxation and
also described the dependence of the soliton velocity on the
relaxation constant $\gamma$.

\bibliography{paper}
\end{document}